# Precise microwave characterization of MgO substrates for HTS circuits with superconducting post dielectric resonator

**Janina Mazierska**[1,2], **Dimitri Ledenyov**[2], **Mohan V Jacob**[2] and **Jerzy Krupka**[3]

[1] Institute of Information Sciences and Technology, Massey University, Palmerston North,
P. Bag 11222, New Zealand
[2] Electrical and Computer Engineering, James Cook University, Townsville, Q4811, Australia
[3] Instytut Mikroelektroniki i Optoelektroniki Politechniki Warszawskiej, Koszykowa 75,
00-662 Warszawa, Poland

**Abstract**
Accurate data of complex permittivity of dielectric substrates are needed for efficient design of HTS microwave planar circuits. We have tested MgO substrates from three different manufacturing batches using a dielectric resonator with superconducting parts recently developed for precise microwave characterization of laminar dielectrics at cryogenic temperatures. The measurement fixture has been fabricated using a $SrLaAlO_3$ post dielectric resonator with $DyBa_2Cu_3O_7$ end plates and silver-plated copper sidewalls to achieve the resolution of loss tangent measurements of $2 \times 10^{-6}$. The tested MgO substrates exhibited the average relative permittivity of 9.63 and $\tan \delta$ from $3.7 \times 10^{-7}$ to $2 \times 10^{-5}$ at frequency of 10.5 GHz in the temperature range from 14 to 80 K.

## 1. Introduction

Magnesium oxide (figure 1(a), [1]) substrates allow for growth of highest quality $YBa_2Cu_3O_7$ and $TlCa_2Ba_2Cu_2O_3$ thin films, despite a certain lattice mismatch. There is also a growing interest in using these substrates for deposition of ferroelectric thin film coatings, crystallization of polymers and plasma display panel technology [2]. The MgO substrates are essentially free of twinning, strain defects and air bubbles; they do not require buffer layers for HTS films; their orientation is typically along the (100) planes.

It is well known that losses of single-crystal dielectric materials may differ significantly. This is due to the fact that even a small number of impurities (which may include Ca, Al, Si, Fe, Cr, B and C in the case of MgO) can change significantly the loss tangent of single crystals. Reported values of $\tan \delta$ of MgO at cryogenic temperatures vary from $2.5 \times 10^{-7}$ to $9 \times 10^{-3}$ as listed in table 1 [2–9]. Quoted values of real relative permittivity $\varepsilon'_r$ also differ, and are in the range from 9 to 9.9. We performed simulations of HTS microstrip resonators on MgO substrate 0.5 mm thick using ENSEMBLE. Assuming $\varepsilon'_r$ of two values, 9.8 and 9.8 + 2%, a surface resistance of HTS film of 350 $\mu\Omega$ and loss tangent of the substrate of $10^{-6}$ we obtained a difference in the computed resonant frequencies of 8.5%. An increase of assumed value of loss tangent from $10^{-6}$ to $10^{-5}$ resulted in a 10% increase in computed insertion loss of the resonator. Hence fabricated HTS filters, designed with the complex permittivity of MgO substrates different from the real value, may exhibit performance unlike that expected.

As is well known, the performance of high temperature superconducting circuits depends on the quality of superconducting thin films (apart from properties of dielectric substrates used). Any irregularity inside the dielectric material or contamination of substrates' surface may result in deterioration of the films' quality and increase of the surface resistance $R_s$. Work recently done at NIST [10] has shown that surface resistance of YBCO films deposited on annealed MgO substrates was smaller than $R_s$ of films on non-annealed substrates (130 m$\Omega$ as compared to 200 m$\Omega$ for temperatures below 40 K). It has also been reported that the third-order intermodulation product of HTS resonators deposited on MgO exhibited a plateau explained as caused by dielectric losses in MgO [5]. It has also been shown that exposure of MgO substrates to environmental conditions or a long term storage in dry $N_2$ before the deposition, as well as lithographic processing, increased very significantly the fraction of the 45° misoriented grains of deposited YBa$_2$Cu$_3$O$_7$ thin films and reduced $J_c$ of YBCO films to 10% of the typical value [11].

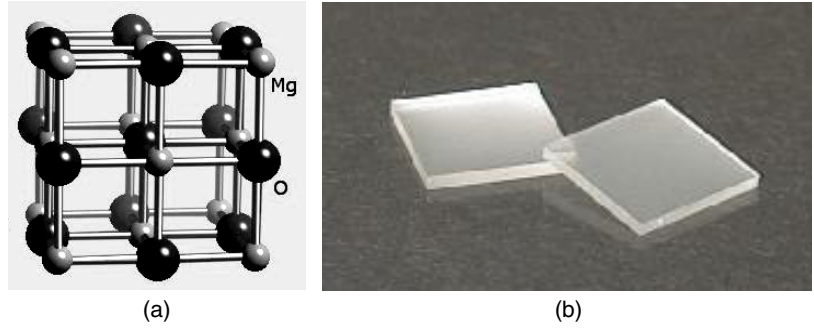

(a)             (b)

**Figure 1.** Magnesium oxide: (a) crystal structure after [1], (b) photograph of substrates after [2].

**Table 1.** Reported values of loss tangent and 'dielectric constant' for MgO.

| Type | Method | Temp. (K) | Freq. (GHz) | $\varepsilon_r$ | $\tan\delta$ | Ref. |
|---|---|---|---|---|---|---|
| Not specified | Not specified | Room temp. | 10 | 9.8 | $9 \times 10^{-3}$ | [2] |
| Bulk | Dielectric resonator | 20–295 | 18 | 9.7–9.8 | $6 \times 10^{-7}$ at 20 K $3.5 \times 10^{-6}$ at 80 K | [3] |
| Bulk | Open ended resonator | 75–280 | 10 | 9 | $2.5 \times 10^{-7}$ at 75 K $4.5 \times 10^{-6}$ at 280 K (7.5 GHz) | [4] |
| Substrate | Stripline resonator | 1.7–20 | 2.3 | Not specified | $2 \times 10^{-5}$ at 1.7 K $1 \times 10^{-6}$ at 20 K | [5] |
| Substrate | Interdigital electrode | 295 | 1 kHz–13 MHz | 9.9 | | [6] |
| Substrate | Near-field microscope | 300 | 0.95 | 9.5 | | [7] |
| Not specified | Not specified | 77, 300 | 10 | 9.6–10 | $6.2 \times 10^{-6}$ at 77 K $2.2 \times 10^{-5}$ at 300 K | [8] |
| Substrate | Not specified | 100 | 300 | 9.65 | $5 \times 10^{-4}$ | [9] |

Screening HTS substrates before the deposition of HTS films by measuring precise values of $\tan\delta$ and $\varepsilon'_r$ could increase the yield of HTS devices and circuits. Hence it may be practical to test all substrates to be used for HTS microwave filters, at a temperature of operation of a system, to eliminate substrates not conforming to the specifications. Such a procedure, although additional, may save time and decrease costs later.

The question arises of whether screening before the deposition process of HTS films (which involves heating to a temperature of 600–800 °C and subsequent cooling to room temperature) is meaningful. If the deposition cycle altered microwave properties of MgO substrates then outcomes of the screening would be of limited value. We performed an investigation into the influence of the annealing process on the complex permittivity of dielectric substrates used for deposition of HTS films [12]. Measurement results of this investigation are still being processed. However, it is feasible to say at this stage that MgO substrates showed negligible change in value of the real part of the relative permittivity and some decrease of loss tangent when exposed to conditions as during the deposition of YBCO films. Hence measurements of the $\varepsilon'_r$ and $\tan\delta$ of MgO substrates before the design of HTS filters will enable more accurate prediction of the resonant frequency of HTS resonators and insertion loss. The screening can also facilitate elimination of substrates with too high losses before the deposition of HTS films.

Methods of measurements of the complex permittivity of HTS substrates are usually based on characterization of single-crystal bulk samples [3, 4, 13] and not actual substrates used for deposition of HTS films. To test the actual substrates the stripline technique is typically used. However, this technique is of limited accuracy and does not allow for usage of substrates after testing. Recently, the split cavity technique [14] and the split post dielectric resonator (SPDR) [15, 16] have been successfully introduced for microwave characterization of low loss single-crystal substrates. The SPDR technique can work at lower frequencies than the split cavity method for smaller complex permittivity dielectrics [17]. However, it requires a very precise centring of the posts and the best loss tangent resolution reported is only $10^{-5}$. A single post fixture (open ended resonator) avoids the need of precise centring and aligning of the posts but was constructed in the past for testing of bulk dielectrics only [18, 13].

Based on the concepts of the single-post resonator for dielectric rods and the split post dielectric resonator for planar dielectrics we have developed a post resonator, as described in [19], for precise measurements of HTS substrates. The resonator was optimized for measurements of low loss planar dielectrics at cryogenic temperatures and was intended for testing HTS substrates before the deposition of HTS thin films. A schematic diagram of the designed resonator is shown in figure 2.

The post resonator contains one SrLaAlO$_3$ dielectric rod resonating at frequency of 10.48 GHz on which a dielectric substrate under test is placed. The post resonator was designed to measure dielectric properties of substrates of thickness of 0.7 mm or smaller, and diameters in the range from 35 to 55 mm. A pair of DyBa$_2$Cu$_3$O$_7$ films was used as the end plates to increase sensitivity of dielectric loss tangent measurements. The constructed resonator was used for precise microwave measurements of $\varepsilon'_r$ and $\tan\delta$ of MgO substrates described below. The TMQF technique [20] to eliminate noise, uncalibrated cables and adaptors and impedance mismatch from the measured data, to ensure high precision in unloaded $Q_0$-factor computations, was used for data processing. A simplified data logging of $S$ parameters was performed as described in [21].

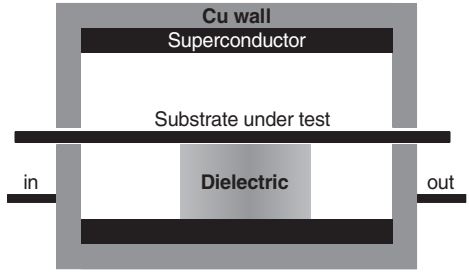

**Figure 2.** Diagram of cryogenic post resonator.

## 2. Characterization of MgO substrates with the post resonator

We have measured nine MgO substrates from three different batches manufactured by Tateho Chemical Industries Co. Ltd, Japan, as a function of temperature from 14 to 84 K. The substrates had the average thickness of 0.508 mm for each batch. Spatial variations in thickness of ±1 μm were observed; measurements of thickness were made with ±1 μm uncertainty. The measurement system based on the superconducting post resonator consisted of a vector network analyser (HP8722C), closed-cycle refrigerator (APD DE204-SL), temperature controller (LTC10), vacuum Dewar and PC. The measurement procedure was the same as described in [19] and for details on the transmission mode $Q$-factor technique readers are referred to [20]. The RF input power level used for measurements was −5 dBm.

The real relative permittivity of the substrates under test has been computed numerically based on the rigorous electromagnetic modelling of the cryogenic post resonant structure using the Rayleigh–Ritz technique [22] as explained in [19]. The maximum absolute uncertainty in $\varepsilon'_r$ resulting from the analysis for our resonator is ±0.2% [23]. Taking into consideration the relative uncertainty of the substrate's thickness we assess the total uncertainty in computations of permittivity to be ±0.5%. Computed values of $\varepsilon'_r$ of the tested MgO substrates are shown in figure 3. The samples exhibited constant values of real relative permittivity when temperature was varied from 15 to 84 K with values of $\varepsilon'_r$ in the range from 9.57 to 9.68 with the maximum difference of 1% (or ±0.5%), hence within the measurement uncertainty. The average $\varepsilon'_r$ value of 9.63 differs from the average of the values reported for the substrates (9.71) by 0.85% and is smaller than highest literature data by 2.8%, confirming existing variations in MgO substrates.

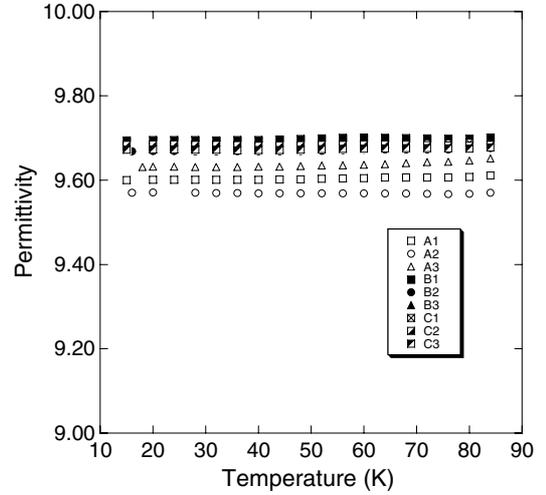

**Figure 3.** Measured $\varepsilon'_r$ of MgO substrates.

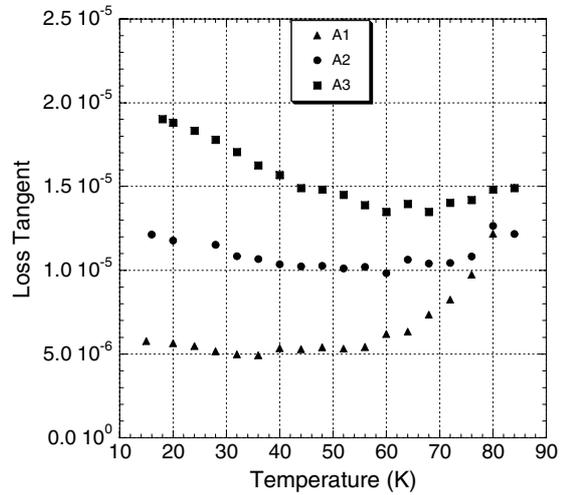

**Figure 4.** Measured loss tangent of batch A.

The loss tangents of the MgO substrates for batches A, B and C at a frequency of 10.48 GHz, computed from the measured unloaded $Q_0$-factors, are given in figures 4–6. The readers are referred to [19] for details of the computation procedure.

In general it can be said that the tested substrates exhibited low values of the loss tangent. The losses of batch A were of the order of $1 \times 10^{-5}$ while for batches B and C losses were below $5 \times 10^{-6}$. The minimum values of $\tan\delta$ for batches A, B and C were $4.93 \times 10^{-6}$, $6.33 \times 10^{-7}$ and $3.8 \times 10^{-7}$ respectively. The temperature dependence of $\tan\delta$ exhibited no evident pattern. However, it could be said that the loss tangent of batch A exhibited a minimum in the temperature dependence. As an example, $\tan\delta$ of sample A1 decreased from $5.79 \times 10^{-6}$ to $4.93 \times 10^{-6}$ at 36 K and then increased to $1.21 \times 10^{-5}$ at 80 K. For sample A1 the minimum occurred at a temperature of 36 K, and A2 and A3 between 60 and 70 K. The samples of batch B showed a more or less similar pattern to A; batch C however exhibited rather maxima versus temperature. The samples from batch A exhibited large variation in losses at lower temperatures (up to four times), much bigger than batches B and C. This is also illustrated in figure 7 where a 'standard deviation' in loss tangent values calculated for the three batches is given. (We use 'SDT' in inverted commas as the number of samples was too small to perform a proper statistical analysis.)

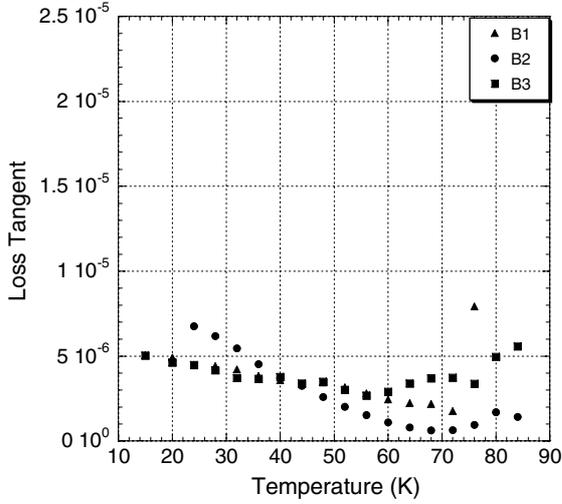

**Figure 5.** Measured tan $\delta$ of batch B.

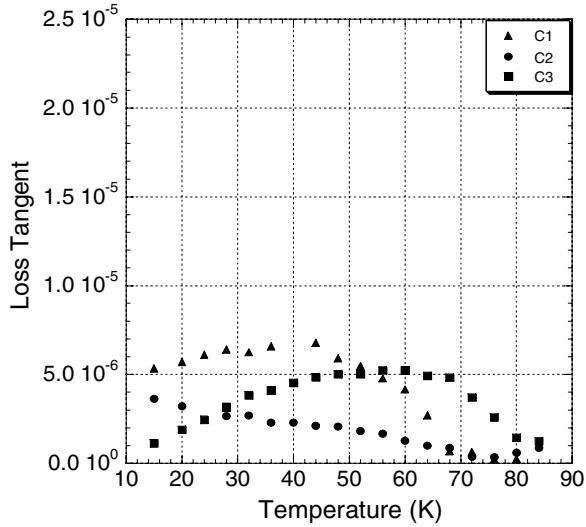

**Figure 6.** Measured tan $\delta$ of batch C.

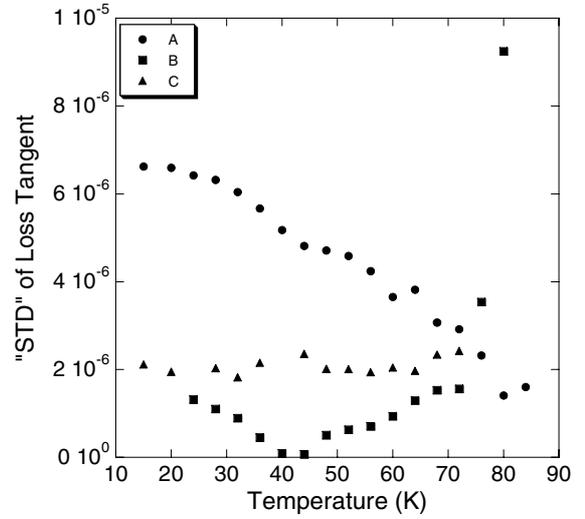

**Figure 7.** 'Standard deviation' in tan $\delta$ values for three MgO batches.

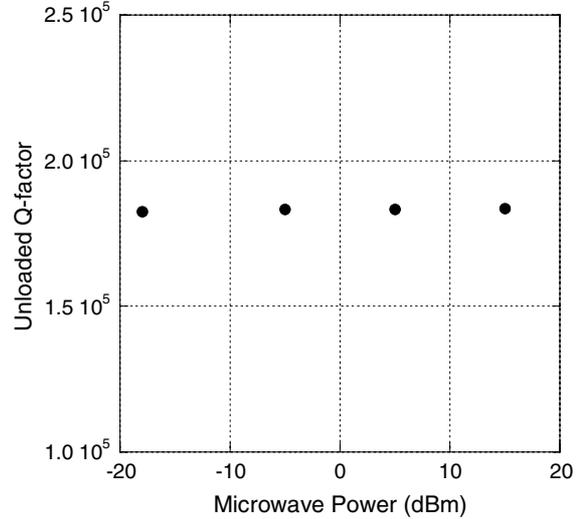

**Figure 8.** $Q_0$-factor versus RF input power at 10.4 GHz and 24 K.

The uncertainty in measured tan $\delta$, ($\Delta_r$ tan $\delta$), values were between 9% and 90% for tan $\delta$ above $2 \times 10^{-6}$ (assuming 1% error in unloaded $Q_0$-factor measurements and 1% uncertainty in the computed energy filling factor for the substrates). The large value of $\Delta_r$ tan $\delta$ is due to the principle of the loss tangent computation based on a difference between the $Q_0$ of the empty resonator and $Q_{0s}$ of the resonator with a substrate under test. For measured values below the resolution of the resonator, $2 \times 10^{-6}$, the uncertainty dramatically increases, even exceeding 500%. Hence in practice for the substrates of batches B and C we assess the upper bound for the losses. A summary comparison of the real relative permittivity, loss tangent, and relative uncertainties of the MgO substrates under test is given in table 2 for temperatures of 15, 52 and 72 K.

We also examined losses of the tested MgO substrates as a function of the input RF power from $-20$ to $+15$ dBm. The obtained dependence of the unloaded $Q_0$-factor of the superconducting post resonator with sample A2 at a temperature of 24 K (presented in figure 8) does not show any nonlinear behaviour with the increased power.

## 3. Conclusions

We have precisely measured the complex permittivity of MgO substrates from differing batches using the recently developed superconducting post resonator. The resonator was optimized to achieve a high resolution and high sensitivity ($Q_0$ of the empty resonator was $2.4 \times 10^5$ at 14 K) to allow for accurate microwave characterization of low loss planar dielectrics. The constructed resonator and developed EM analysis enabled computations of the real part of relative permittivity with uncertainty of $\pm 0.5\%$ (allowing for $\pm 0.3\%$ uncertainty in thickness of substrates). The tested MgO substrates exhibited constant values of $\varepsilon'_r$ with temperature in the range from 14 to 84 K. The average value of $\varepsilon'_r$ was 9.63 with variation of $\pm 0.5\%$ between the samples, equal to the measurement uncertainty. The average $\varepsilon'_r$ differs by 2.8% as compared to the highest literature data. The differences of 0.5% and 2.8% in $\varepsilon'_r$ would result in a change of $f_{res}$ of an HTS resonator deposited on this substrate from a predicted value of 1.850 to 1.845 GHz and 1.826 GHz respectively.

**Table 2.** Measured parameters of MgO substrates at 15, 52 and 72 K.

| | Substrate | 15 K | 52 K | 72 K |
|---|---|---|---|---|
| Real relative permittivity | A1 | $9.60 \pm 0.5\%$ | $9.60 \pm 0.5\%$ | $9.61 \pm 0.5\%$ |
| | A2 | $9.57 \pm 0.5\%$ | $9.57 \pm 0.5\%$ | $9.57 \pm 0.5\%$ |
| | A3 | $9.63 \pm 0.5\%$ | $9.63 \pm 0.5\%$ | $9.64 \pm 0.5\%$ |
| | B1 | $9.69 \pm 0.5\%$ | $9.70 \pm 0.5\%$ | $9.70 \pm 0.5\%$ |
| | B2 | $9.67 \pm 0.5\%$ | $9.67 \pm 0.5\%$ | $9.67 \pm 0.5\%$ |
| | B3 | $9.68 \pm 0.5\%$ | $9.68 \pm 0.5\%$ | $9.69 \pm 0.5\%$ |
| | C1 | $9.68 \pm 0.5\%$ | $9.68 \pm 0.5\%$ | $9.69 \pm 0.5\%$ |
| | C2 | $9.68 \pm 0.5\%$ | $9.68 \pm 0.5\%$ | $9.68 \pm 0.5\%$ |
| | C3 | $9.67 \pm 0.5\%$ | $9.67 \pm 0.5\%$ | $9.68 \pm 0.5\%$ |
| $\tan \delta$ | A1 | $5.79 \times 10^{-6} \pm 27\%$ | $5.32 \times 10^{-6} \pm 35\%$ | $8.24 \times 10^{-6} \pm 30\%$ |
| | A2 | $1.20 \times 10^{-5} \pm 14\%$ | $1.01 \times 10^{-5} \pm 19\%$ | $1.04 \times 10^{-5} \pm 24\%$ |
| | A3 | $1.90 \times 10^{-5} \pm 9\%$ | $1.45 \times 10^{-5} \pm 14\%$ | $1.40 \times 10^{-5} \pm 18\%$ |
| | B1 | $5.07 \times 10^{-6} \pm 30\%$ | $3.17 \times 10^{-6} \pm 58\%$ | $1.77 \times 10^{-6} \pm 134\%$ |
| | B2 | | $2.02 \times 10^{-6} \pm 91\%$ | $6.5 \times 10^{-7} \pm 360\%$ |
| | B3 | $5.03 \times 10^{-6} \pm 30\%$ | $3.02 \times 10^{-6} \pm 61\%$ | $3.74 \times 10^{-6} \pm 64\%$ |
| | C1 | $5.33 \times 10^{-6} \pm 29\%$ | $5.48 \times 10^{-6} \pm 34\%$ | $6.5 \times 10^{-7} \pm 331\%$ |
| | C2 | $3.63 \times 10^{-6} \pm 42\%$ | $1.8 \times 10^{-6} \pm 101\%$ | $3.7 \times 10^{-7} \pm 564\%$ |
| | C3 | $1.13 \times 10^{-6} \pm 131\%$ | $5.0 \times 10^{-6} \pm 37\%$ | $3.6 \times 10^{-6} \pm 60\%$ |

The average value of loss tangent of the substrates measured at 10.48 GHz from 14 to 84 K was approximately $5 \times 10^{-6}$. Measured losses differ between the samples, as well as their temperature dependences. Losses of batch A (with the highest values of $\tan \delta$) exhibited a minimum at a temperature between 50 and 60 K. The other two batches (of very low loss) showed differing temperature dependences: batch B exhibited minima, the other one maxima. The measurement uncertainty in loss tangent values bigger than $5 \times 10^{-6}$ was less than 30% and for $\tan \delta$ larger than $2 \times 10^{-6}$ it was smaller than 91% (estimating the relative uncertainty in the $Q_0$-factor of $\pm 1\%$). For substrates of $\tan \delta$ of order $10^{-7}$ the uncertainty of measurements is very high and only an upper bound for loss can be assessed. The tested MgO substrates did not show RF power dependence up to 15 dBm.

The developed cryogenic post resonator has proved to be a useful tool for precise selection of MgO substrates prior to deposition of HTS films to allow for precise design of HTS passive circuits for applications in wireless base station filters.

## Acknowledgments

This work was partly supported by the ARC Large Grant A00105170 (James Cook University) and the Linkage Grant LX0242351. MVJ acknowledges JCU CRIG support. The authors would like to thank Mr Kazuhisa Niwano from Tateho Chemical Industries Co., Ltd for MgO samples.